\documentclass[12 pt]{article}
\usepackage{times}
\usepackage{subfigure}

\usepackage{pifont}
\usepackage{floatrow}
\usepackage{caption}
\usepackage{mathrsfs}
\usepackage[fleqn]{amsmath}
\usepackage{amsfonts,amsthm,amssymb,mathrsfs,bbding}
\usepackage{txfonts}
\usepackage{graphics,multicol}
\usepackage{graphicx}
\usepackage{color}
\usepackage{caption}
\usepackage{indentfirst}
\usepackage{cite}
\usepackage{latexsym,bm}
\usepackage{enumerate}
\usepackage{epsfig}
\evensidemargin=\oddsidemargin\topmargin=-1.5cm

\theoremstyle{definition}

\addtocounter{section}{0}

\numberwithin{equation}{section}

\let\trueiiint=\iiint
\def\iiint{\mathop{\textstyle\trueiiint}\limits}
\def\intinfty{\int\limits_{\!\!-\infty\,\,}^{\,\,\infty\!\!}\kern-0.0em}
\def\iintinfty{\mathop{\int\!\!\int}\limits_{\!\!-\infty\,\,}^{\,\,\infty\!\!}\kern-0.0em}
\def\iiintinfty{\mathop{\int\!\!\int\!\!\int}\limits_{\!\!-\infty\,\,}^{\,\,\infty\!\!}\kern-0.0em}
\def\Real{{\mathbb R}}

\let\<=\langle
\let\>=\rangle
\let\^=\hat
\def\~#1{{\-ox{\sf#1}}}
\def\N{{\mathbb N}}

\def\bbeta{{\bm{\theta}}}

\def\btheta{{\bm{\theta}}}
\let\-=\mathbf

\def\@#1{{\cal #1}}

\def\N{{\mathcal{N}}}
\def\Cov{\mathrm{Cov}}

\let\<=\langle
\let\>=\rangle

\let\^=\hat

\def\circ{\ifmmode\mathchar"220E\else$\mathchar"220E$\fi}
\def\@#1{{\cal #1}}
\def\x{{\bm{x}}}
\def\y{{\bm{y}}}

\def\r{\bm{r}}

\begin{document}
\title{Inferring the unknown parameters in Differential Equation by Gaussian Process Regression with Constraint}

\author{Y. Zhou$^{\ddag}$
\quad {and} \quad
H. Wang${^{\ddag} }^*$ \\[2mm]
{\it $^{\ddag}$School of Mathematics and Statistics} \\
{\it Central South University} \\
{\it Changsha 410083, P.R. China}
}


\date{}

\maketitle {\flushleft\large\bf Abstract }

Differential Equation (DE) is a commonly used modeling method in various scientific subjects such as finance and biology. 
The parameters in DE models often have interesting scientific interpretations, but their values are often unknown and need to be estimated from the measurements of the DE. 
In this work, we propose a Bayesian inference framework to solve the problem of estimating the parameters of the DE model, from the given noisy and scarce observations of the solution only.
A key issue in this problem is to robustly estimate the derivatives of a function from noisy observations of only the function values at given location points, under the assumption of a physical model in the form of differential equation
governing the function and its derivatives.
To address the key issue, we use the Gaussian Process Regression with Constraint (GPRC) method which jointly model the solution, the derivatives, and the parametric differential equation, to estimate the solution and its derivatives.
For nonlinear differential equations, a Picard-iteration-like  approximation of linearization method is used so that the GPRC can be still iteratively applicable.
A new potential which combines the data and equation information, is proposed and used in the likelihood for our inference.
With numerical examples, we illustrate that the proposed method has competitive performance against existing approaches for estimating the unknown parameters in DEs.

\begin{flushleft}
\textbf{Keywords:} Gaussian process,
Bayesian inference,
Inverse problems
\end{flushleft}

\section{Introduction}\label{s:intro} 

Differential equations (DEs) which include ordinary differential equation (ODE) and partial differential equation (PDE), are used to model a wide variety of physical phenomena. 
ODEs describe the dynamics of continuously changing processes by relating a process and its rate of change in biomedical research and other scientific areas.  
PDE models are commonly used to model complex dynamic system in applied science such as physics, biology and finance.
The forms of DE models are usually proposed by experts based on their prior knowledge and understanding of the dynamic system. 
But the values of the parameters in DE models are often unknown and need to be estimated from noisy measurements of the dynamics system.
Because most DEs have no analytic solutions and can only be solved with numerical method. 
Traditional methods for estimating DE parameters require repeatedly solving DEs numerically under thousands of candidate parameter values. 

A lot of statistical methods have been developed to estimate the parameters in DE modes by repeatedly solving the DE models numerically, which could be time consuming.
Specifically, a series of method in the study of HIV dynamics \cite{ho1995rapid}\cite{wei1995viral}\cite{wu1999population}\cite{wu2005statistical}\cite{wu1998estimation} and some hierarchical Bayesian methods \cite{putter2002bayesian} \cite{huang2006hierarchical} are proposed with use of numerical ODE solver.
As for PDEs, Muller and Timmer \cite{muller2002fitting} solved the target least-squares type minimization problem using an extended multiple shooting method.
It requires careful parameterization of the initial condition and then numerically solves the PDEs. 
This leads to a pretty high computational load and there is growing need in developing efficient estimation methods for DE models.

Another strategy to estimate parameters of DE is the two-stage method, which in the first stage estimates the solutions and its derivatives from noisy observations using data smoothing methods without considering differential equation models, and then in the second stage estimates of the DE parameters are obtained by least squares. 
Liang and Wu \cite{liang2008parameter} developed a two-stage method for a general first-order ODE model, using local polynomial regression in the first stage, and estimated asymptotic properties of the estimator. 
Similarly, Chen and Wu \cite{chen2008efficient} developed local estimation for time-varying coefficients.
Bar, Hegger, and Kantz \cite{bar1999fitting} modeled unknown PDEs using multivariate polynomials of sufficiently high order, and the best fit was chosen by minimizing the least squares error of the polynomial approximation.
Based on the estimated functions, the PDE parameters were estimated using least squares \cite{muller2004parameter}. 
The issues of noise level and data resolution were addressed extensively in this approach. See also Parlitz and Merkwirth \cite{parlitz2000prediction} and Voss et al. \cite{voss1999amplitude} for more example. 

As mentioned before, a straightforward two-stage strategy, though easy to implement, has difficulty in estimating derivatives of the dynamic process accurately, leading to biased estimates of the DE parameter.
There are some advanced methods\cite{ramsay2007parameter}\cite{cao2011robust}\cite{cao2012penalized}, which approximate the solution and its derivatives by using basis-function with both considering noisy observations and equation constraint. 
Then parameters are estimated by optimizing certain criteria. 
Especially, Xiaolei Xun\cite{xun2013parameter} propose two joint modeling schemes: (a) a parameter cascading or penalized profile likelihood approach and (b) a fully Bayesian treatment.
This type methods usually have some sensitive hyper-parameters, which needed be carefully design and may have the parameter estimates converge to a local minima, otherwise global optimization is computationally intensive. In addition, The ability of basis functions to fit complex solution function and its derivatives is also a challenge.

Here we propose using a new statistical approach which  is based on Gaussian process, called Gaussian process regression with constraint (GPRC) method \cite{wang2020explicit}, to 
accurately and robustly predict the solution and  derivatives appearing in the differential equations. 
GPRC can efficiently make the predictions with both considering the observations and DE constraint in a joint Bayesian framework.
Compared with basis-function expansion method, the Gaussian process (GP) models \cite{seeger2004gaussian} provide a class of flexible and efficient  non-parametric fits 
both for fitting noisy data and solving noisy DEs.  
The prior knowledge can also be easily encoded by the covariance of the GP.
To achieve parameter estimation, we construct a likelihood by employing a new potential term which contains the DE  and data residuals and the Metropolis-Hastings(MH) algorithm is used to infer the unknown parameters in Bayes' rule.

The structure of the paper is organized as follows: 
We first review the problem setup in Section \ref{set:Problemsetup}. 
The review of GPRC method for estimating the solution and its derivatives is presented in Section \ref{set:method}. Specificly, the Picard iterative linearization for nonlinear DE is carefully formulated in \ref{set:picard}, based on the last example in \cite{wang2020explicit}.
Bayesian inference framework is introduced in Section \ref{set:inference}, where we propose a new potential of the likelihood.
Numerical examples are presented in Section \ref{set:numerical}, to demonstrate the effectiveness of the proposed method compared with two-stage method. Finally, Section \ref{set:conclusion} offers some closing remarks.
\section{Problem setup}
\label{set:Problemsetup}

Inferring the unknown parameters in differential equation is a classical problem in the inverse problem. Differential equations refer to the  relationship between the solution function and its derivatives. We consider a multidimensional dynamic process, $u(\x)$ , where $\x = (x_1, \dots, x_D)^T\in \mathbb{R}^D$ is a multidimensional argument. 
The general formula for differential equation can be expressed as
\begin{equation}
\label{eq:pde}
\mathcal{F}(\x, u,\frac{\partial u}{\partial x_1},\dots, \frac{\partial u}{\partial x_p},\dots, \frac{\partial^2 u} {\partial x_i\partial x_j},\dots; \bbeta) = 0,
\end{equation}
 where $\bbeta=(\theta_1,\dots,\theta_q)$ is the  vector of  unknown parameters. In the following, we denote $\mathcal{F}(\x, u(\x), \dots; \bbeta)$ by the short-hand notation $\mathcal{F}(u(\bm{x}); \bbeta)$. The left-hand side of Eq. \eqref{eq:pde} consists of the solution $u(\x)$ and its derivative terms ($\frac{\partial u} {\partial x_i}, \frac{\partial^2 u} {\partial x_i\partial x_i}, \dots, $). 
In practice , we may not   directly observe $u(\x)$, but we  can observe the noisy measurement  $y(\x)$. For simplicity, we assume that there is a measurement error between $u(\x)$ and $y(\x)$, the noisy measurements satisfy 
\begin{equation} 
\label{y}
y_i = u(\x_i) + \epsilon_i,
\end{equation}
where $i = 1,\dots, n$, is the observation index and $\epsilon_i$ is the measurement error. We assume this measurement error follows an independent and identically distributed Gaussian distribution with zero mean and variance $\sigma_{obs}^2$.

In practice, parameters $\bbeta$ are often unknown and need to be determined. Our objective is to estimate the parameters from the limited noisy data $\{\x_i, y(\x_i)\}_{i=1, \dots, n}$.

\section{GPRC}
\label{set:method}
For estimating the unknown parameter $\btheta$, accurate estimations of the solution and its derivatives are necessary and helpful. 
To infer a nonlinear function from its noisy measurements at a given set of inputs is a classical statistical learning problem and a vast of well-established methods, ranging from polynomial and spline to kernel method and neural network, are available for many important applications.
However, if the interest is the derivatives and there is no additional observations of derivatives, the problemof estimating derivatives is more challenging and subtle than estimating the function itself and less attentions have been given to this issue.
In this section, we briefly introduce an estimator GPRC \cite{wang2020explicit}, which has shown the ability of accurately estimating the derivatives in ordinary differential equation and partial differential equation problems and here is used to estimate the solution and its derivatives with a guess parameter $\btheta$. 
\subsection{ GPRC in linear differential equation}
\label{subset:gprc}
Specifically, the solution $u(\x)$ is assumed as a Gaussian process with zero mean and the covariance function $k_{uu}(\x,\x',\bm{\gamma})$, which is denoted as 
\begin{equation}
\label{eq:GPprior}
u(\x) \sim \mathcal{GP}(0, k_{uu}(\x,\x';\bm{\gamma})),
\end{equation}
where, $\bm{\gamma}$ denotes the hyper-parameters of the kernel function. One property of the Gaussian process in our favor is that any linear transformation, such as differentiation and integration, of a Gaussian process is still a Gaussian process. With the assumption \eqref{eq:GPprior}, we consider the case when $\mathcal{F}(\cdot; \bbeta)$ is a linear operator or a combination of different linear operators, denoted as $\mathcal{L}^\btheta$ with given $\bbeta$. We introduce a random function $r(\x)$ as the residual of the linear differential equation $\mathcal{L}^\btheta u = 0$ for convenience:
\begin{equation*}
r(\x):=\mathcal{L}^\btheta u(x).
\end{equation*} 
Then the function $r(\x)$ is also a mean-zero Gaussian process 
\begin{equation*}
r(\x) = \mathcal{GP}(0, k_{rr}(\x,\x';\bm{\gamma})),
\end{equation*}
where $k_{rr}(\x,\x') = \text{Cov}(\mathcal{L}^\btheta u(\x), \mathcal{L}^\btheta u(\x'))$ denotes the covariance function of $\mathcal{L}^\btheta u$ between $\x$ and $\x'$. The following fundamental relationship between the kernels $k_{uu}$ and $k_{rr}$, $k_{ur}$, $k_{ru}$ is well-known (see  e.g. \cite{seeger2004gaussian,theore2003Solving}),\
\begin{equation}
\label{eq:cov_ru}
\begin{split}
k_{rr}(\x,\x';\bm{\gamma}) &= \mathcal{L}_{\x}^\btheta \mathcal{L}_{\x'}^\btheta k_{uu}(\x,\x';\bm{\gamma}),\\
k_{ur}(\x,\x';\bm{\gamma}) &= \mathcal{L}_{\x'}^\btheta k_{uu}(\x,\x';\bm{\gamma}),\\
k_{ru}(\x,\x';\bm{\gamma}) &= \mathcal{L}^\btheta_{\x}k_{uu}(\x,\x';\bm{\gamma}),
\end{split}
\end{equation}
Based on the Gaussian assumption and the covariance expressions between $u$ and $r$ in \eqref{eq:cov_ru}, a joint inference framework of Gaussian process regression for the available observation data of $u$ and $r$ can be naturally constructed.
By interpreting the equation $r := \mathcal{L}^\btheta u = 0$ as the constraint of the function $r$, GPRC will significantly improve the accuracy of estimation of solution and its derivatives due to the additional equation information.

\textbf{Training a Gaussian Process}

The training process is to find the optimal parameters $\bm{\beta} = \{\bm \gamma, \sigma_{obs}^2\}$ by maximum likelihood estimation (MLE). 
Given $n$  noisy observations of the state $u$  
at  $n$ points  $\{\x_i: 1\leq i\leq n\}$,  we  denote  the state vector $\y \equiv [y_1, y_2,\dots,y_n]^T\in \Real^{n\times 1}$, the residual vector $\r \equiv [r_1, r_2,\dots,r_n]^T\in \Real^{n\times 1}$ and 
the training point matrix $X \equiv [\x_1, \x_2,\dots,\x_n]^T\in \Real^{n\times D}$.
Let   $Y = \begin{bmatrix} \y\\ \r\end{bmatrix} \in \Real^{2n}$. 
Then
 the   negative log marginal likelihood
of  $p(Y\vert \bm{\beta})$ 
 has the following expression
\begin{equation}
\label{eq:logmarginal}
\begin{split}
-\log p(Y| \bm{\beta}) = \frac{1}{2}\log (\det{K})+\frac{1}{2}Y^T K^{-1}Y + \frac{n}{2}\log 2\pi,
\end{split}
\end{equation}
where  the $2n \times 2n$ matrix $K$ is defined by
\begin{equation*}
K = \begin{bmatrix}
&K_{uu}+ \sigma^2_u I & K_{ur}\\
&K_{ru} & K_{rr} + \sigma^2_r I
\end{bmatrix},
\end{equation*}
 The matrices $K_{uu}, K_{ur}$,  $K_{ru}$ and $K_{rr}$
 correspond to the kernel functions $k_{uu}, k_{ur}, k_{ru}$ and $k_{rr}$ respectively in \eqref{eq:GPprior} and \eqref{eq:cov_ru}, evaluated at the $n$ points $X $.  \\
The minimisation of the negative log marginal likelihood  is a non-convex optimisation task. However gradients are easily obtained
\begin{equation}
\label{eq:partialderivative}
\begin{split}
\frac{\partial L}{\partial \beta_k}=\frac{1}{2}tr(K\frac{\partial K^{-1}}{\partial \beta_k})+\frac{1}{2}Y^T\frac{\partial K}{\partial \beta_k}K^{-1}\frac{\partial K}{\partial \beta_k} Y,
\end{split}
\end{equation}
where $\beta_k $  indicates the $i$th hyper-parameter of $\bm{\beta}$ and therefore standard gradient optimisers can be used to minimize the negative log marginal likelihood \eqref{eq:logmarginal} within the gradient \eqref{eq:partialderivative}. 

\textbf{Prediction}
After the hyper-parameters in the GPRC are computed,
the prediction of the function $u(\x)$ or its derivatives of interest,
denoted as a function $l(\x)$ (e.g., $l(\x)=\partial_{\x} u(\x)$),
at a new  test point $\x_*$ is described  below.
The covariance function of the GP $l(\x)$ is denoted by $k_{ll}(\x,\x')$ by the convenience.

With a given   point $\x_*$,  the differential equation provides  the fact $r(\x_*)=0$, which
is a useful observation to incorporate the Bayesian inference.
To enhance this condition locally, $m$ equally-spaced points around $\x_*$, which are referred as the extended set $\chi:= \{ \x_*^j: 1\le j\le m  \}$ are actually considered in the prediction.
The set $\chi$ is similar to the window/scale  concept in local polynomial regression and thus adaptive strategy is possible.
Therefore, the value of the equation constraints in the extended set should be zero, i.e., $\r_{\chi}=\bm 0$. 
These points $\x_*^j$ are supposed to resolve a characteristic length for the residual process $r(\x)$. 

This is the posterior distribution for a specific set of test case,   
\begin{align}
\label{eq:poster_u}
&\bar{u}(\x_*) = K_{u_*\bullet}\widehat{K}_{u\r_{\chi}}^{-1}\begin{bmatrix}\y\\\r_{\chi}\end{bmatrix},
S_u(\x_*)=K_{uu}(\x_*,\x_*) - K_{u_*\bullet} \widehat{K}_{u\r_{\chi}}^{-1}K_{u_*\bullet}^T.\\
\label{eq:poster_l}
&\bar{l}(\x_*) = K_{l_*\bullet}\widehat{K}_{u\r_{\chi}}^{-1}\begin{bmatrix}\y\\\r_{\chi}\end{bmatrix},
S_l(\x_*)=K_{ll}(\x_*,\x_*) - K_{l_*\bullet}\widehat{K}_{u\r_{\chi}}^{-1}K_{l_*\bullet}^T,
\end{align}
where 
$$\widehat{K}_{u\r_{\chi}}=\begin{bmatrix}
K_u&K_{u \r_{\chi}}\\
K_{u \r_{\chi}}^T  &K_{\r_{\chi}}
\end{bmatrix}\in \Real^{(n+m)\times (n+m)},$$
$K_u = K_{uu} + \sigma^2_u \bm{I}_u\in \Real^{n\times n}$,
 $ K_{\r_{\chi}} = K_{\r_{\chi}\r_{\chi}}+ \sigma^2_r \bm{I}_{r} \in \Real^{m\times m}$,
  $K_{u_*} = k_{uu}(\x_*,\x_*)$,
$K_{l_*}= k_{ll}(\x_*,\x_*)$
and $K_{u_*\bullet} = [K_{u_*u},  K_{u_*\r_{\chi}}]\in \Real^{1\times (n+m)}$ and $K_{l_*\bullet} = [K_{l_*u} ,K_{l_*\r_{\chi}}]\in \Real^{1\times (n+m)}$.  The posterior variances $S_u(\x_*)$ and $S_l(\x_*)$ can be used as good indicators of how confident the predictions are.

\subsection{Picard iterative linearization for nonlinear differential equation}
\label{set:picard}                                                                                                             
For nonlinear differential equations, the equation constraint can't be formulated as a Gaussian process since the product of Gaussian processes is not a Gaussian process anymore. 
Here an iterative linearization method is introduced for nonlinear equations, motivated by the Picard iteration method\cite{coddington1955theory} , which is an extension of method used in Van der Pol equation example in \cite{wang2020explicit}. 

Assume there have an initial guess of the solution $\hat{u}_0$ and its  derivative $\hat{l}^i_0$ which can be fitted by any classical method, like Gaussian process regression and basis function expansion. 
Here the hat notation $\hat{\cdot}$ represents an estimation and we are only interest in certain derivative terms which is useful for lineariztion. 
So the $i$th derivative term $\hat{l}^i$ is used to represent all derivatives of interest for simplicity. 
Then any nonlinear term of the nonlinear equation, which is composed of $K$ solution or its derivative terms, can be simply approximated by replacing $K-1$ terms with the corresponding given initial guess of the solution $\hat{u}_0$ and its derivatives $\hat{l}^i_0$, and only keep one term. 
This is a quite straightforward linearlization strategy and can be easily applied. 
There is a trick that replacing the solution or low order derivatives, would be preferred than the one who replacing the high order derivatives of the nonlinear term in the approximation. 
That is because the instability of numerical differentiation of the fitted function solely from the data which is the measurement of solution with adding noise, will be more serious in estimating higher order derivative. The linearization of $\mathcal{F}(u(\bm{x}); \bbeta)$ is expressed as 
\begin{equation}
\label{eq:linear}
\mathcal{F}(u, l^i,\dots;  \hat{u}_0, \hat{l}^i_0,\dots; \bbeta) = 0.
\end{equation}
Based on linearization in the Eq. \eqref{eq:linear}, GPRC introduced in Section \ref{subset:gprc} can be used to estimate a new solution $\hat{u}$ and derivative $\hat{l}^i$. 
Then an iterative scheme can be constructed with the new pair $\{\hat{u}, \hat{l}^i\}$. 
Roughly speaking in the $n$th cycle, given the current guess of solution $\hat{u}_n$ and derivatives $\hat{l}^i_n$, we can compute a new (and possibly better) approximation $\{\hat{u}_{n+1}, \hat{l}^i_{n+1} \}$ using
\begin{equation}
\label{eq:picard_iteration}
\{\hat{u}_{n+1}, \hat{l}^i_{n+1} \}= \textbf{GPRC}(\mathcal{F}(u, l^i, \dots; \hat{u}_n, \hat{l}^i_n,\dots;\bbeta), \-y),
\end{equation}
where $\textbf{GPRC}(\cdot)$ indicates the estimator of GPRC whose output is the posterior mean functions of the solution and its derivatives. 
Our stopping criterion is the sum of RMSE values of solution and the RMSE values of all derivatives of interest, which is expressed by 
\begin{equation*}
e = \int (||\hat{u}_{n+1} - \hat{u}_{n}||^2 + \sum_i ||\hat{l}^i_{n+1} - \hat{l}^i_{n}||^2)p(\x) d\x,
\end{equation*}
where $|| \cdot||^2$ denotes the L2 normalization. If $e$ is small than a give small positive threshold $\epsilon_{rmse}$,  then we think the iteration reaches stability.

\section{Inference}
\label{set:inference}
We are interest in the problem of estimating the unknown parameters $\bm{\theta}$ from the indirect observations $\y$  and the DE constraint. So for inferring the unknown parameters, both data equations \eqref{y} and DE constraint equation \eqref{eq:pde} should be satisfied.

In Bayesian setting, the prior belief about the parameter $\bm{\theta}$ is encoded in the probability distribution $\pi_{prior}(\bbeta)$. Here we use a uniform distribution as prior for highlighting the action of likelihood function. Our aim is to infer the distribution of $\bbeta$ conditioned on the given data $\bm{y}$, i.e., the posterior distribution $\pi(\bbeta|\-y)$. By the Bayes' rule, we have 
\begin{equation}
\label{eq:posterior_mcmc}
\pi(\bbeta|\-y)\propto \exp(-\alpha*\eta(\bbeta;\-y;\mathcal{F}(\cdot)))*\pi_{prior}(\bbeta),
\end{equation}
where $\alpha$ is a scale hyperparameter, which is used to adjust the likelihood value to a reasonable scale, and $\eta(\bbeta;\-y;\mathcal{F}(\cdot))$ is called potential of the likelihood $\exp(-\alpha*\eta(\bbeta;\-y;\mathcal{F}(\cdot)))$, specific
\begin{equation}
\label{eq:potential}
\eta(\bbeta;\-y;\mathcal{F}(\cdot)) = \frac{1}{2}||\-y -  \hat{u}(X;\bbeta)||^2 +\frac{1}{2}\int [\mathcal{F}(\hat{u}(\-x;\bbeta);\bbeta)]^2 p(\bm{x})d\bm{x}.
\end{equation}
Fidelity to the PDE can be measured by $\int [\mathcal{F}(\hat{u}(\-x;\bbeta);\bbeta)]^2 p(\bm{x})d\bm{x}$, while fidelity to the data can be measured by 
$||\-y -  \hat{u}(X;\bbeta)||^2$.  Equal weights for data and equation information are assumed in E.q \eqref{eq:potential}.

The unnormalized posterior \eqref{eq:posterior_mcmc} can be easily sampled using MCMC method such as  Metropolis-Hastings (MH) algorithm \cite{chib1995understanding}, Gibbs sampling \cite{gilks1992adaptive} and DRAM \cite{haario2006dram} et al.
Here we use MH sampling algorithm to draw samples from the unnormalized posterior \eqref{eq:posterior_mcmc}.
The Metropolis-Hastings (MH) algorithm is one of the most popular MCMC methods. 
An MH step of invariant distribution $\pi(\btheta)$ and proposal distribution  $ \pi_q(\btheta^{i+1} |\btheta^i) $ involves sampling a candidate value $\btheta^{i+1}$ given the current value $\btheta^i$  according to $ \pi_q(\btheta^{i+1}| \btheta^i)$. 
The Markov chain then moves towards $\btheta^{i+1}$ with acceptance probability $\mathcal{A}(\bbeta^{(i)},\bm{\theta}^{i+1})=\min\{1,\frac{\pi(\bbeta^{i+1}|\-y)q(\bbeta^{(i)}|\bbeta^{i+1})}{\pi(\bbeta^{(i)}|\-y)q(\bbeta^{i+1}|\bbeta^{(i)})}\}$, otherwise it remains at $\btheta^i$.
In our work, we draw from $\pi(\bbeta|\-y)$ using MH sampling:
\begin{enumerate}
\item Initialise $\bm{\theta}^{(0)}$
\item For $i$ = 0 to $N$ -1\\ 
(a) Sample $\btheta^{i+1}$ from the proposal distribution $\pi_q(\bm{\theta}^{i+1}|\bm{\theta}^{(i)})$.\\
(b) Compute $\mathcal{A}(\bbeta^{(i)},\bm{\theta}^{i+1})=\min\{1,\frac{\pi(\bbeta^{i+1}|\-y)\pi_q(\bbeta^{(i)}|\bbeta^{i+1})}{\pi(\bbeta^{(i)}|\-y) \pi_q(\bbeta^{i+1}|\bbeta^{(i)})}\}$ and sample $u$ from $\mathcal{U}[0,1]$.\\
(c) If $u < \mathcal{A}(\bbeta^{(i)},\bm{\theta}^{i+1})$ then accept $\bbeta^{i+1}$.  Otherwise, set $\bbeta^{(i+1)}=\bbeta^{(i)}$.
\end{enumerate}

\section{Numerical implementation}
The GPRC method can be used for estimating the solution and derivatives in either ODE or PDE problems \cite{wang2020explicit}. 
Here we investigate the parameter estimation problem in three scenarios: linear ODE, nonlinear ODE and nonlinear PDE, to verify the effectiveness of our method and make a comparison with two-stage method.
It's worth noting that only one iteration is applied in Picard linearization procedure in Eq. \eqref{eq:picard_iteration} because we only replace the solution $u$ in nonlinear terms that is always estimated well directly from data.
In two-stage method, we also use Gaussian process regression as statistical model to approximate the state and derivatives from the observations \cite{rai2019gaussian}. 
Same Gaussian process based regression model, used in two-stage method and our method can eliminate the error caused by the model difference.
In two-stage method, the data information in Eq. \eqref{eq:potential} is fixed due to the solution and its derivative estimations would not vary with different $\bm{\theta}$.
Then the unknown parameters can be detected by only minimizing the parametric equation constraint residual. 
So the potential is reduced to  
\begin{equation*}
\eta(\bbeta;\mathcal{F}(\cdot)) = \int [\mathcal{F}(\hat{u}(\-x;\bbeta);\bbeta)]^2 p(\bm{x})d\bm{x}.
\end{equation*} 
The same radial basis function (RBF) kernel are shared in both GPRC and GPR (used in two-stage method),
\begin{equation*}
k_{uu}(\x, \x',\bm{\gamma}) = \gamma^2_\alpha \exp(-\frac{1}{2}\sum_{d=1}^D\gamma_d(\x_d-\x_{d}')^2).
\end{equation*} 
with hyper-parameter $\gamma^2_\alpha$ and $\gamma_d$. 
The kernel $k_{lu}(\x, \x',\bm{\gamma})$ can be computed based on \eqref{eq:cov_ru}. For instance the kernel of first order derivative term $\frac{du(t)}{d t}$ in ODE can be expressed as:
\begin{equation*}
k_{lu}(t, t',\bm{\gamma}) = \frac{k_{uu}(t, t',\bm{\gamma})}{d t}=-k_{uu}(t,t',\bm{\gamma})\gamma_d(t-t').
\end{equation*}
The expression of other kernel for a general differential operator can be computed similarly.
\label{set:numerical}
\subsection{$1$D toy example}
\begin{figure}[htpb]
\centering
\includegraphics[width=0.48\linewidth]{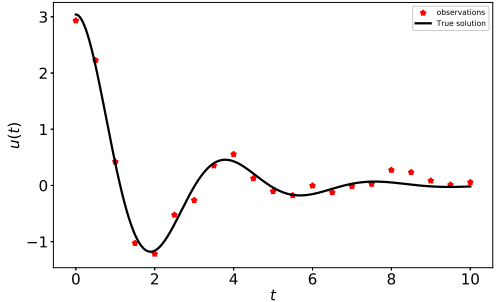}
\caption{Illustration of the data and the true solution}
\label{fig:linearode}
\end{figure}

First we consider the following  $1$-dimension linear ordinary differential equation,
\begin{equation}
\label{eq:exam1ode}
u''(t)+\theta_1u'(t)+\theta_2u(t)=0.
\end{equation}
where $u$ is a function of time $t$, $'$ and $''$ denotes first  and second order derivative operators, respectively. 
$\btheta = [\theta_1, \theta_2]$ is the unknown parameters which need to refer. 
Our true parameter values are taken as $\theta^*_1=1$,$\theta^*_2=3$. Then $21$ observations are evenly measured in the time interval $[0,10]$ with Gaussian noise $\mathcal{N}(0, \sigma^2_{obs}=0.1)$.  The observations and the true solution $u$ are shown in Fig. \ref{fig:linearode}.
We assume the prior of the sate function $u$ is a zero-mean Gaussian process expressed in \eqref{eq:GPprior}. 
As discussed in Section \ref{set:method}, the equation constraint $r = u'' + \theta_1 u' + \theta_2u$ is also a Gaussian process, $
r\sim \mathcal{GP}(0,k_{rr}(\x,\x';\bm{\gamma})).$
With the property of covariance, the covariance between $r$ and $u$ can be expanded as
\begin{align*}
\Cov(r,u) &=\Cov(u'' + \theta_1u' + \theta_2u,u)\\
&=\Cov(u'',u) + \theta_1\Cov(u',u)+\theta_2\Cov(u,u).
\end{align*}
So,  $k_{ru} = k_{u''u} + \theta_1k_{u'u}+\theta_2k_{u,u}$.
 Similarly, the kernel functions 
 corresponding to covariances $
   \Cov(r,u'), \Cov(r,u'')$ and $\Cov(r,r)$ are expressed as 
 \begin{align*}
 k_{ru'} &= k_{u''u'} + \theta_1k_{u'u'}+\theta_2k_{u,u'},\\
 k_{ru''} &= k_{u''u''} + \theta_1k_{u'u''}+\theta_2k_{u,u''},\\
 k_{rr}& = k_{u'' u''} + {\theta_1}^2 k_{u'u'} + {\theta_2}^2  k_{uu} + 2{\theta_1} k_{u''u'}+2{\theta_2} k_{u''u}+ 2{\theta_1}{\theta_2} k_{u'u}.
\end{align*}  

\begin{figure}[htpb]
\centering
\includegraphics[width=0.48\linewidth]{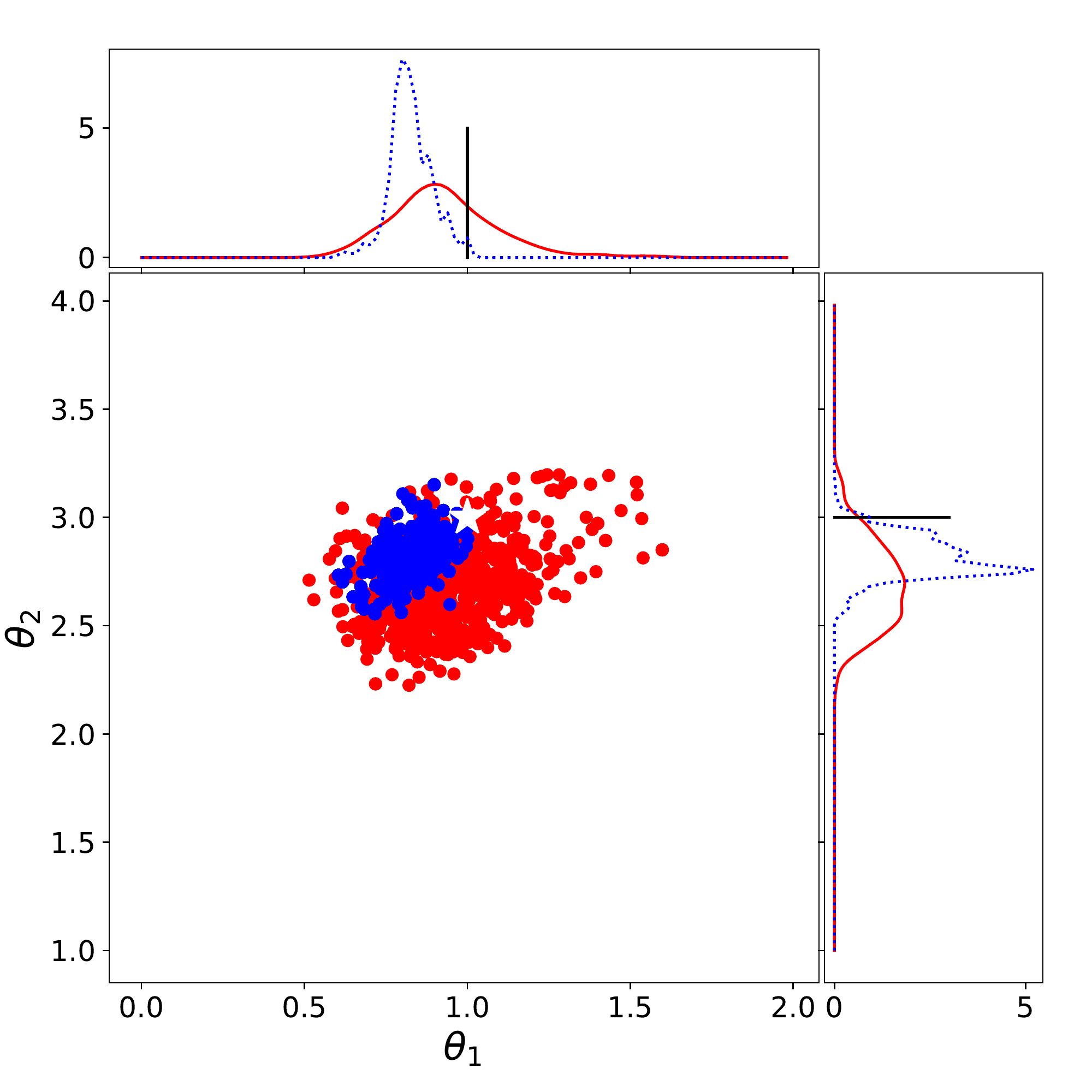}
\includegraphics[width=0.48\linewidth]{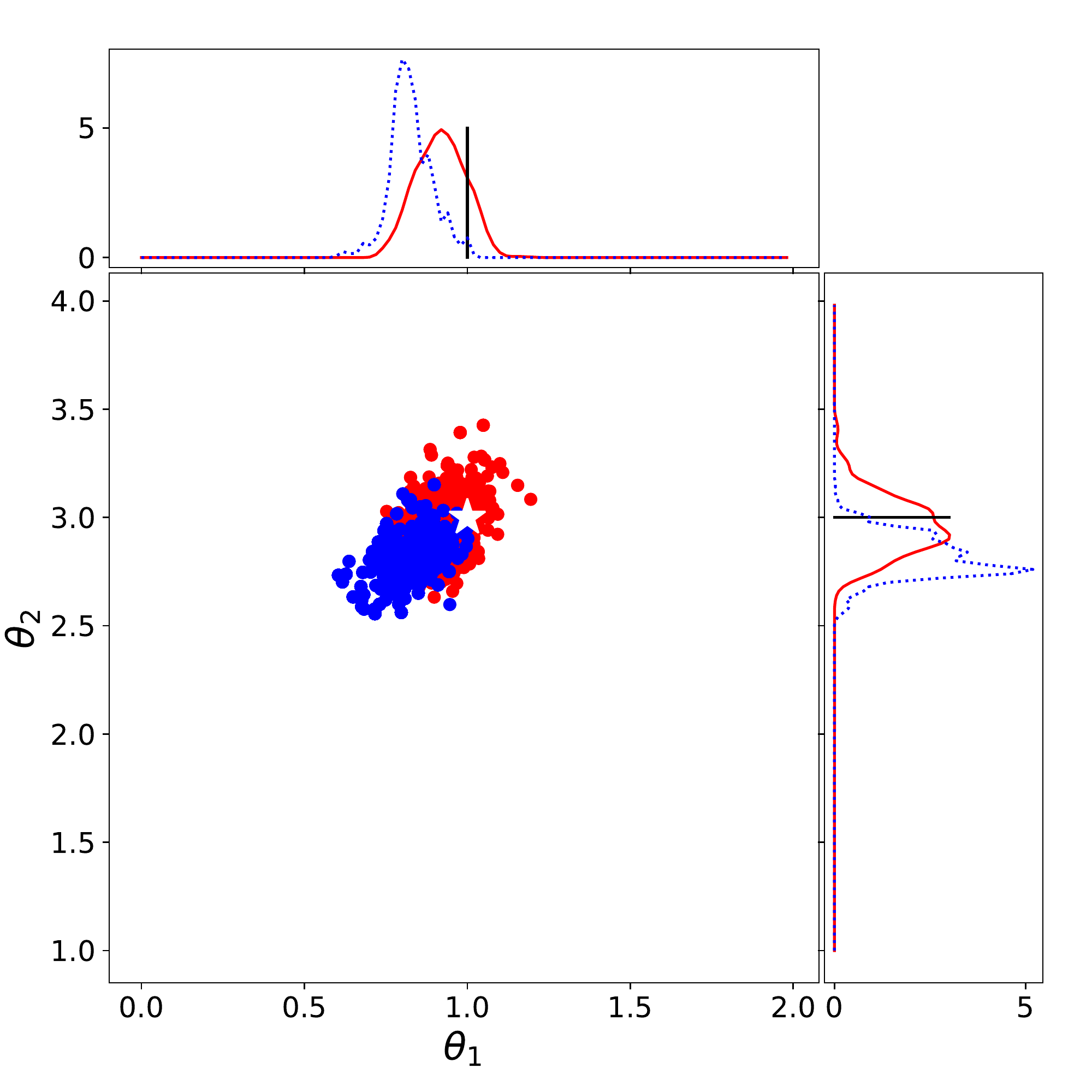}
\caption{Posterior samples and marginal pdfs, computed by our method (red) and two-stage method (blue) in different noise levels (Left: $\sigma^2_{obs}=0.1$; and Right: $\sigma^2_{obs}=0.05$). Posterior samples are shown in the scatter diagram and true parameter (black line) and two marginal pdfs are drawn in subplots corresponding to different noise levels. White star is our true parameter location;}
\label{fig:linearode_posterior}
\end{figure}

By equation \eqref{eq:poster_u} and \eqref{eq:poster_l},
 the posterior distributions $p(u_*|\y, \r_{\chi}) = \N(m_u, \Sigma_u)$, $p(u_*'|\y, \r_{\chi}) = \N(m_{u'}, \Sigma_{u'})$ and $p(u_*''|\y, \r_{\chi}) = \N(m_{u''}, \Sigma_{u''}) $ are  given below
\begin{align*}
m_{u_*} &= K_{\bullet u_*}^T\widehat{K}_{u\r_{\chi}}^{-1}Y,
\quad 
\Sigma_{u_*} =K_{u_*} - K_{\bullet u_*}^T\widehat{K}_{u\r_{\chi}}^{-1}K_{\bullet u_*}\\
m_{u'_*} &= K_{\bullet u_*'}^T\widehat{K}_{u\r_{\chi}}^{-1}Y,
\quad
\Sigma_{u'_*} = K_{u'_*} - K_{\bullet u_*'}^T\widehat{K}_{u\r_{\chi}}^{-1}K_{\bullet u_*'}\\
m_{u''_*} &= K_{\bullet u_*''}^T\widehat{K}_{u\r_{\chi}}^{-1}Y,
\quad
\Sigma_{u''_*} = K_{u''_*} - K_{\bullet u_*''}^T\widehat{K}_{u\r_{\chi}} ^{-1}K_{\bullet u_*''},
\end{align*}
where   $Y = [\y, \r_{\chi}]^T$.
This is  the posterior estimation of state and derivative functions.
In GPRC algorithm, we set $\sigma_r^2= 10$  and $\chi= [t_*]$, which means only thee predicted point is included in the extend set $\chi$. Although we would sacrifice some precision of prediction due to the small extend set, this setting can efficiently speed up the calculation. 
Then we can obtain the posterior estimation of solution and its derivatives within any given $\btheta$. 
A Gaussian proposal distribution $q(\btheta^*|\btheta^{(i)})=\mathcal{N}(\btheta^{(i)},0.6^2)$ is used as the proposal distribution in our MH sampler with $\alpha = 100$ in our likelihood function \eqref{eq:posterior_mcmc}. 
$5000$ samples are drawn from the posterior. 
As a comparison same number of samples are obtained by two-stage method and the both posterior results are shown in  Figure \ref{fig:linearode_posterior}.
Obviously from the figure on the left, we can find that the samples (red scattered points) computed by our method have a wider variance, which means a bigger uncertainty of the posterior is estimated. 
In the marginal pdf figures, the first dimension marginal pdf value at $\theta_1^*$ computed by our method (red line) are larger than one computed by two-stage method (blue dash)  and the second dimension marginal pdf values computed by two method at $\theta_1^*$ are equal. These results indicate the posterior estimation computed by our method is more credible. 
This point would be better clear in a small-noise case because the bias caused by the numerical error of derivatives estimation would be highlights when the measure noise is small.

We then consider the small-noise case where we use the same implementation configurations as in the large-noise case. 
We change the noise level from $0.1$ to $0.05$ and collect the same number of observations at same locations. 
The samples from posterior are shown in Figure \ref{fig:linearode_posterior} (right).
Similar to the large-noise case, the figure shows that the results of the two-stage are of low accuracy, while our method yield rather good results.

\subsection{Van der Pol equation}
\begin{figure}[htpb]
\centering
\includegraphics[width=0.45\linewidth]{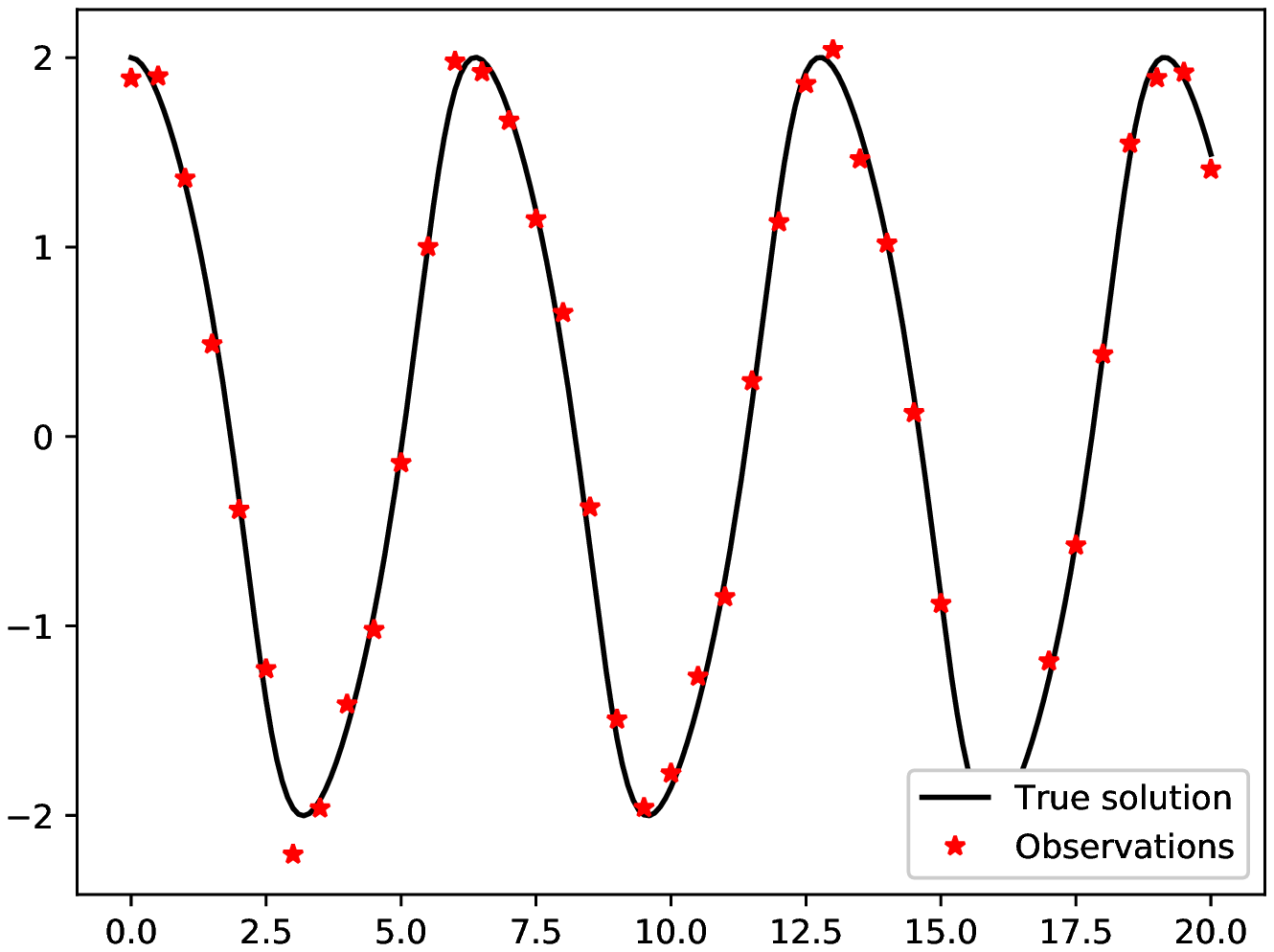}
\includegraphics[width=0.5\linewidth]{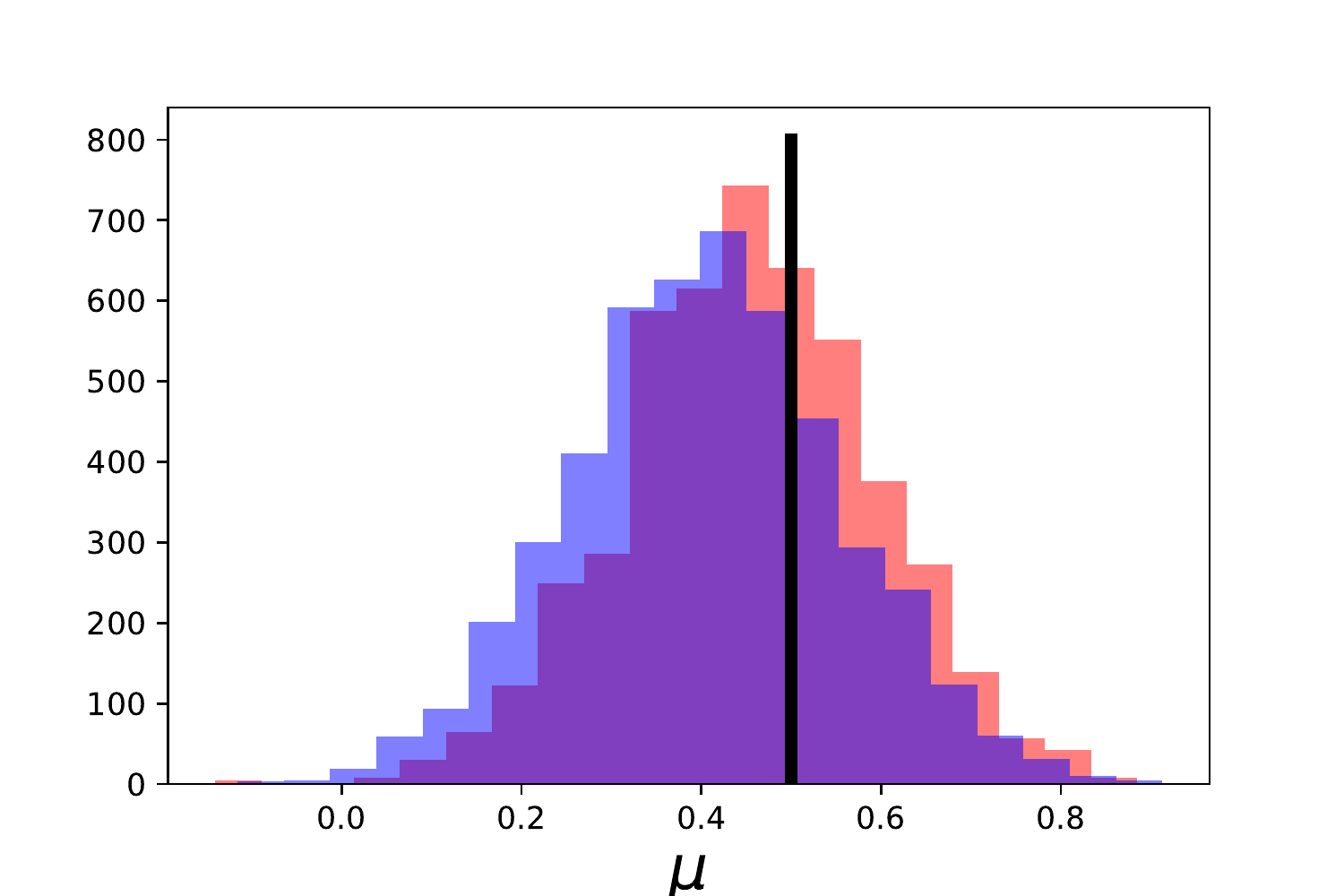}
\caption{Left: the solution $u$ (black line) with true parameter $\mu^*=0.5$ and our noisy observations (red star); Right: histograms of the posterior samples computed by our method (red) and two-stage method (blue), separately. Black line is our true parameter $\mu^*=0.5$; }
\label{fig:vdp_equation}
\end{figure}

The Van der Pol model  is described by 1-dimensional nonlinear ODE equation,
\begin{equation}
\label{eq:vande}
u''- \mu(1-u^2)u' + u=0,
\end{equation}
where $\mu$ is the unknown parameter. For this nonlinear differential equation, we first use the GPR method to obtain initial guess of the  solution $ \hat{u}_0$ only from data. 
Then we can obtain a linearizaiton form of \eqref{eq:vande}, by using the approximation $\hat{u}_0$ to replace the original $u$ of nonlinear term, expressed as 
\begin{equation*}
u''+\mu(1-\hat{u}_0^2)u'+u=0.
\end{equation*}
In this example 41 observations are measured at every $0.5$ time units on the interval [0,20] with given true parameter $\mu^*=0.5$. 
The clean data are corrupted with additive Gaussian noise with variance $\sigma_{obs}^2=0.01 $. 
The figure on the left in Fig \ref{fig:vdp_equation} shows the true solution $u$ and the observations.
In GPRC,  $\sigma_r^2=0.1$ is set and $10$ points which are evenly selected in the domain $[t_*-1, t_*+1]$, form the extend set $\chi$ .  In the MH sampling procedure, the  design of the proposal distribution $q(\theta^*|\theta^{(i)})$ is $\mathcal{N}(\theta^{(i)},0.6^2)$. 
The histogram of the posterior samples computed by both our method and two-stage method are shown in the Fig. \ref{fig:vdp_equation} (right). 
In the figure, the true parameter (black line) are closer to the mean of red histogram of samples computed by our method than the one of blue histogram, computed by two-stage method.  

\subsection{The KdV Equation}
\begin{figure}[htpb]
\centering
\includegraphics[width=0.48\linewidth]{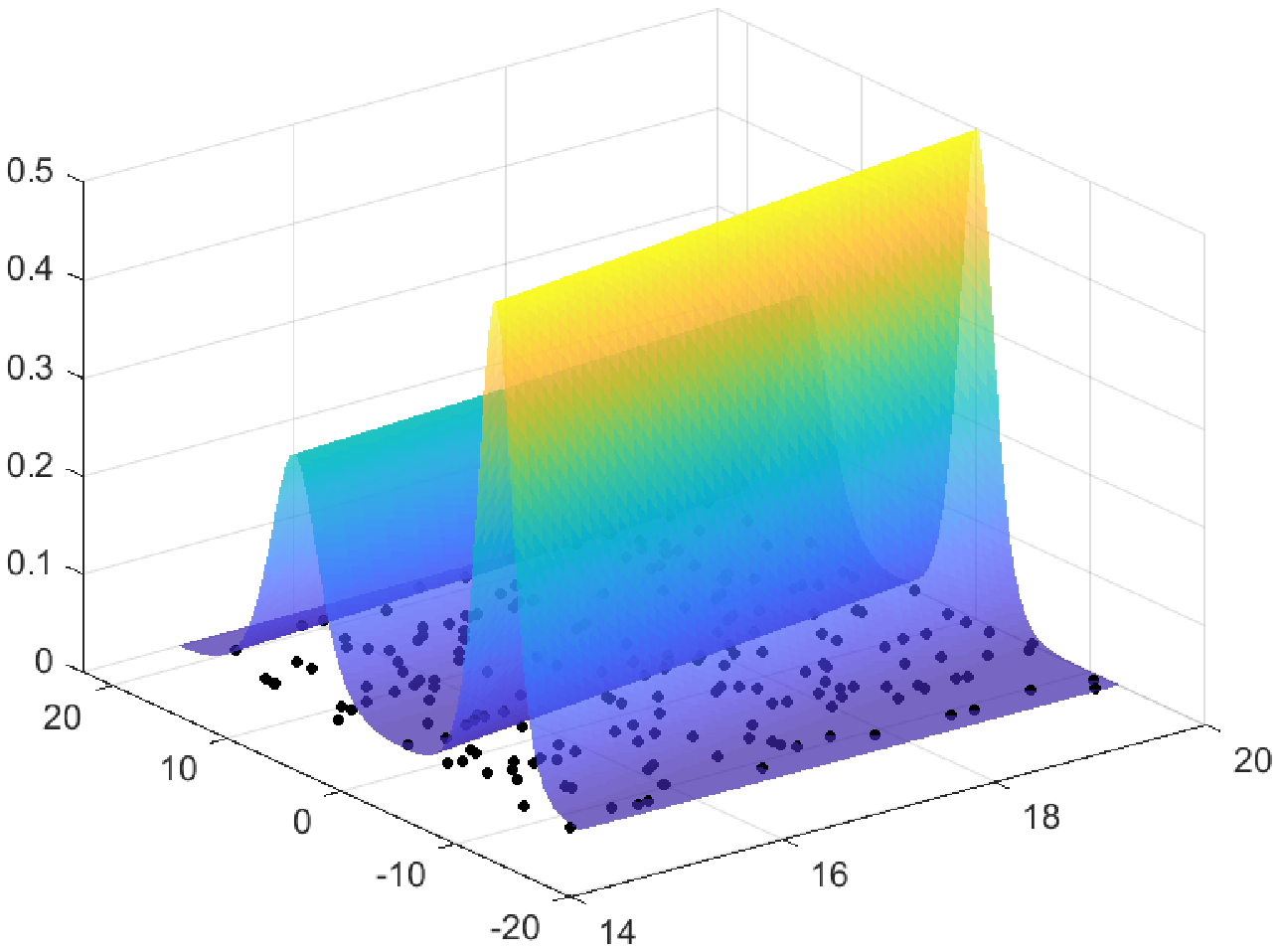}
\includegraphics[width=0.45\linewidth]{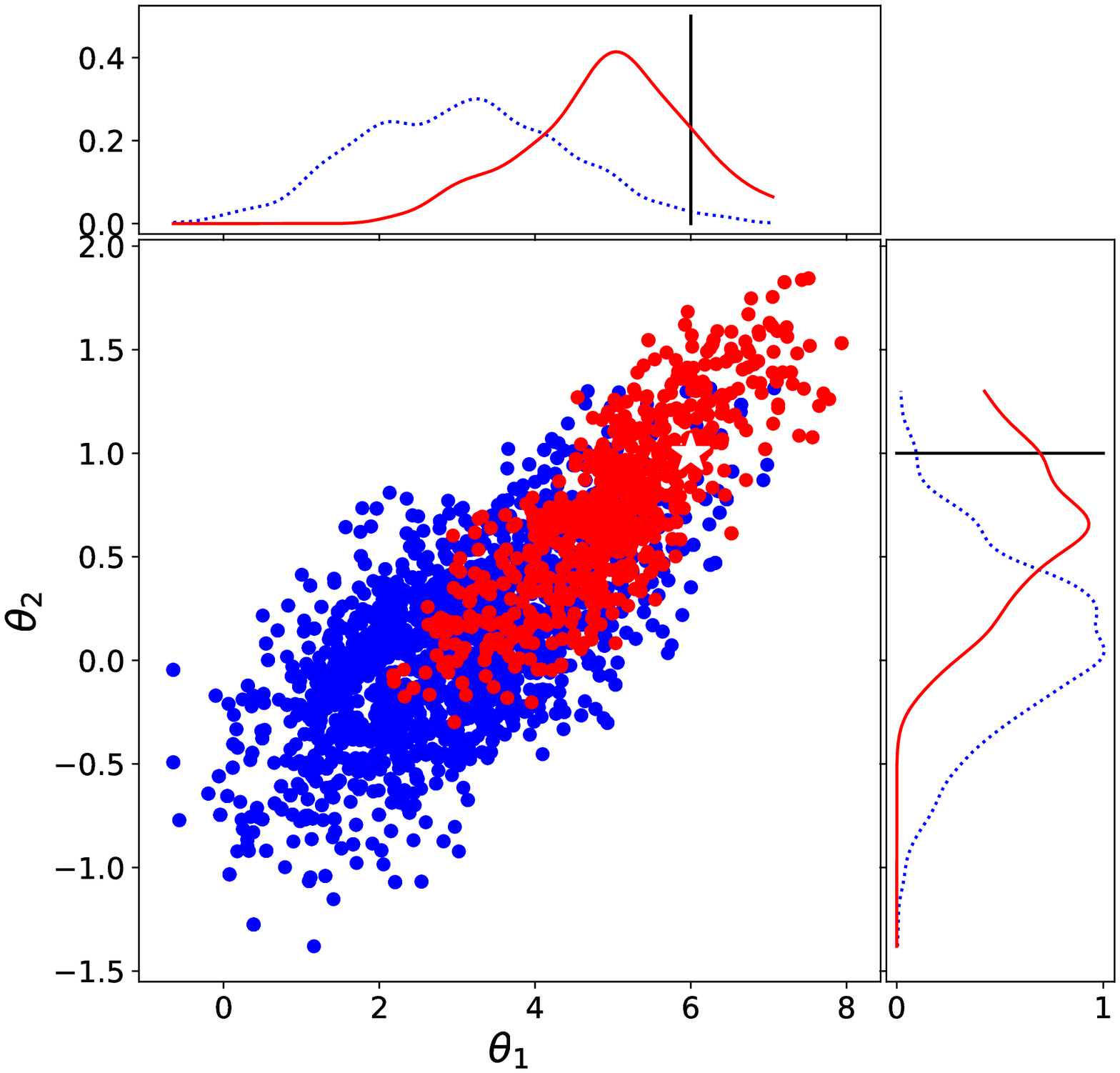}
\caption{Left: the illustration of true solution and the observations (black scattered points); Right: posterior samples (scattered points) and two corresponding marginal pdfs computed by our method ( red) and two-stage method ( blue). White star is our true parameter;}
\label{fig:example3_post}
\end{figure}

The KdV equation is a nonlinear, dispersive partial differential equation for function $u$ of two real variables, space $x\in \mathbb{R}^{1\times1}$ and time $t$:
\begin{equation*}
\frac{\partial u}{\partial t} + \theta_1 u \frac{\partial u}{\partial x} + \theta_2 \frac{\partial^3 u}{\partial x^3} = 0,
\end{equation*} 
It is an asymptotic simplification of Euler equations used to model waves in shallow water. 
It can also be viewed as Burger's equation with added dispersive term. 
In this equation two parameters need to be estimated from some noisy observations of function $u$. The true solution is created using a spectral method with $512$ spatial points and $200$ timesteps. More details about this model refer \cite{rudy2017data}. 
This is a highly non-linear PDE. So $200$ solution values are randomly selected from the given discrete solution in the time domain $[15, 20]$ and spatial domain $[-21, 26]$, within our true parameters are $\theta^*_1=6$ and $\theta^*_2=1$. 
A Gaussian noise is added on the selected solution values with zero mean and $0.1$ variance.
The true solution and the observations are shown in the Fig. \ref{fig:example3_post} (left). 

In MH sampler procedure, $5000$ samples are obtained by the given proposal distribution $\mathcal{N}(0, 0.6I)$.  The posterior samples and marginal distributions are shown in right figure of Fig. \ref{fig:example3_post}.
As shown, the true parameter (white star) is closer to the center of the posterior samples computed by our method than the one of two-stage method. In addition, at the true parameters $\theta^*_1$ and $\theta^*_2$, the values of marginal PDF computed by our method are larger than the one computed by two-stage method, express that our method yields a better posterior estimation.

\section{Conclusion}
\label{set:conclusion}
In conclusion, we make use of a powerful tool: the Gaussian Process Regression with Constraint method (GPRC) to the problem of parameter estimation. 
We propose a new potential function which combines the data information and equation information, by using the accurate estimation of solution and its derivatives computed by GPRC. 
Then posterior samples are drawn from the product of prior and likelihood with this type potential by MH sampler.
A two-stage method are also illustrated in each example for comparison.
With both linear and nonlinear examples, we illustrate that the proposed method can be a competitive tool to estimate the unknown differential equation parameters.
Finally we discuss some limitations of the proposed method as well as some issues that need to be address in the future. 
Firstly for any given parameters, the GPRC would estimate the solution and derivatives by a training procedure which is time-costing.
Maybe an efficient surrogate can be constructed for the likelihood and  an adaptive procedure\cite{wang2018adaptive} can be employed for efficient sampling in future.
Secondly, the observations with large bias would lead to ineffectiveness of our method. 
GPRC would reduce the numerical error of derivative operator by considering the DE information, but the biased observations probably let it fail.












\bibliographystyle{plain}

\bibliography{References}

\begin{thebibliography}{10}

\bibitem{2003An}
Christophe Andrieu, Nando~De Freitas, Arnaud Doucet, and Michael~I. Jordan.
\newblock An introduction to mcmc for machine learning.
\newblock {\em Machine Learning}, 50(1/2):p.5--43, 2003.

\bibitem{bar1999fitting}
Markus B{\"a}r, Rainer Hegger, and Holger Kantz.
\newblock Fitting partial differential equations to space-time dynamics.
\newblock {\em Physical Review E}, 59(1):337, 1999.

\bibitem{cao2011robust}
J~Cao, L~Wang, and J~Xu.
\newblock Robust estimation for ordinary differential equation models.
\newblock {\em Biometrics}, 67(4):1305--1313, 2011.

\bibitem{cao2012penalized}
Jiguo Cao, Jianhua~Z Huang, and Hulin Wu.
\newblock Penalized nonlinear least squares estimation of time-varying
  parameters in ordinary differential equations.
\newblock {\em Journal of computational and graphical statistics},
  21(1):42--56, 2012.

\bibitem{chen2008efficient}
Jianwei Chen and Hulin Wu.
\newblock Efficient local estimation for time-varying coefficients in
  deterministic dynamic models with applications to hiv-1 dynamics.
\newblock {\em Journal of the American Statistical Association},
  103(481):369--384, 2008.

\bibitem{chib1995understanding}
Siddhartha Chib and Edward Greenberg.
\newblock Understanding the metropolis-hastings algorithm.
\newblock {\em The american statistician}, 49(4):327--335, 1995.

\bibitem{coddington1955theory}
Earl~A Coddington and Norman Levinson.
\newblock {\em Theory of ordinary differential equations}.
\newblock Tata McGraw-Hill Education, 1955.

\bibitem{gilks1992adaptive}
Walter~R Gilks and Pascal Wild.
\newblock Adaptive rejection sampling for gibbs sampling.
\newblock {\em Journal of the Royal Statistical Society: Series C (Applied
  Statistics)}, 41(2):337--348, 1992.

\bibitem{theore2003Solving}
Thore Graepel.
\newblock Solving noisy linear operator equations by gaussian process:
  Application to ordinary and partial differential equations.
\newblock In {\em International Conference on Machine Learning}, pages
  234--241, 2003.

\bibitem{haario2006dram}
Heikki Haario, Marko Laine, Antonietta Mira, and Eero Saksman.
\newblock Dram: efficient adaptive mcmc.
\newblock {\em Statistics and computing}, 16(4):339--354, 2006.

\bibitem{ho1995rapid}
David~D Ho, Avidan~U Neumann, Alan~S Perelson, Wen Chen, John~M Leonard, and
  Martin Markowitz.
\newblock Rapid turnover of plasma virions and cd4 lymphocytes in hiv-1
  infection.
\newblock {\em Nature}, 373(6510):123--126, 1995.

\bibitem{huang2006hierarchical}
Yangxin Huang, Dacheng Liu, and Hulin Wu.
\newblock Hierarchical bayesian methods for estimation of parameters in a
  longitudinal hiv dynamic system.
\newblock {\em Biometrics}, 62(2):413--423, 2006.

\bibitem{liang2008parameter}
Hua Liang and Hulin Wu.
\newblock Parameter estimation for differential equation models using a
  framework of measurement error in regression models.
\newblock {\em Journal of the American Statistical Association},
  103(484):1570--1583, 2008.

\bibitem{muller2004parameter}
TG~M{\"u}ller and Jens Timmer.
\newblock Parameter identification techniques for partial differential
  equations.
\newblock {\em International Journal of Bifurcation and Chaos},
  14(06):2053--2060, 2004.

\bibitem{muller2002fitting}
Thorsten~G M{\"u}ller and Jens Timmer.
\newblock Fitting parameters in partial differential equations from partially
  observed noisy data.
\newblock {\em Physica D: Nonlinear Phenomena}, 171(1-2):1--7, 2002.

\bibitem{parlitz2000prediction}
Ulrich Parlitz and Christian Merkwirth.
\newblock Prediction of spatiotemporal time series based on reconstructed local
  states.
\newblock {\em Physical review letters}, 84(9):1890, 2000.

\bibitem{putter2002bayesian}
Hein Putter, SH~Heisterkamp, JMA Lange, and F~De~Wolf.
\newblock A bayesian approach to parameter estimation in hiv dynamical models.
\newblock {\em Statistics in medicine}, 21(15):2199--2214, 2002.

\bibitem{rai2019gaussian}
Pankaj~Kumar Rai and Shivam Tripathi.
\newblock Gaussian process for estimating parameters of partial differential
  equations and its application to the richards equation.
\newblock {\em Stochastic Environmental Research and Risk Assessment},
  33(8-9):1629--1649, 2019.

\bibitem{ramsay2007parameter}
Jim~O Ramsay, Giles Hooker, David Campbell, and Jiguo Cao.
\newblock Parameter estimation for differential equations: a generalized
  smoothing approach.
\newblock {\em Journal of the Royal Statistical Society: Series B (Statistical
  Methodology)}, 69(5):741--796, 2007.

\bibitem{rudy2017data}
Samuel~H Rudy, Steven~L Brunton, Joshua~L Proctor, and J~Nathan Kutz.
\newblock Data-driven discovery of partial differential equations.
\newblock {\em Science Advances}, 3(4):e1602614, 2017.

\bibitem{seeger2004gaussian}
Matthias Seeger.
\newblock Gaussian processes for machine learning.
\newblock {\em International journal of neural systems}, 14(02):69--106, 2004.

\bibitem{voss1999amplitude}
Henning~U Voss, Paul Kolodner, Markus Abel, and J{\"u}rgen Kurths.
\newblock Amplitude equations from spatiotemporal binary-fluid convection data.
\newblock {\em Physical review letters}, 83(17):3422, 1999.

\bibitem{wang2018adaptive}
Hongqiao Wang and Jinglai Li.
\newblock Adaptive gaussian process approximation for bayesian inference with
  expensive likelihood functions.
\newblock {\em Neural computation}, 30(11):3072--3094, 2018.

\bibitem{wang2020explicit}
Hongqiao Wang and Xiang Zhou.
\newblock Explicit estimation of derivatives from data and differential
  equations by gaussian process regression.
\newblock {\em arXiv preprint arXiv:2004.05796}, 2020.

\bibitem{wei1995viral}
Xiping Wei, Sajal~K Ghosh, Maria~E Taylor, Victoria~A Johnson, Emilio~A Emini,
  Paul Deutsch, Jeffrey~D Lifson, Sebastian Bonhoeffer, Martin~A Nowak,
  Beatrice~H Hahn, et~al.
\newblock Viral dynamics in human immunodeficiency virus type 1 infection.
\newblock {\em Nature}, 373(6510):117--122, 1995.

\bibitem{wu2005statistical}
Hulin Wu.
\newblock Statistical methods for hiv dynamic studies in aids clinical trials.
\newblock {\em Statistical methods in medical research}, 14(2):171--192, 2005.

\bibitem{wu1999population}
Hulin Wu and A~Adam Ding.
\newblock Population hiv-1 dynamics in vivo: applicable models and inferential
  tools for virological data from aids clinical trials.
\newblock {\em Biometrics}, 55(2):410--418, 1999.

\bibitem{wu1998estimation}
Hulin Wu, A~Adam Ding, and Victor De~Gruttola.
\newblock Estimation of hiv dynamic parameters.
\newblock {\em Statistics in medicine}, 17(21):2463--2485, 1998.

\bibitem{xun2013parameter}
Xiaolei Xun, Jiguo Cao, Bani Mallick, Arnab Maity, and Raymond~J Carroll.
\newblock Parameter estimation of partial differential equation models.
\newblock {\em Journal of the American Statistical Association},
  108(503):1009--1020, 2013.

\end{thebibliography}

\end{document}